\documentclass{IEEEtran4PSCC}
\usepackage{cite}
\ifCLASSINFOpdf
   \usepackage[pdftex]{graphicx}
   \graphicspath{ {./images/} }
\else
   \usepackage[dvips]{graphicx}
\fi
\usepackage{xcolor}

\usepackage{hyperref}
\hypersetup{
colorlinks=true,       % false: boxed links; true: colored links
linkcolor=black,          % color of internal links
citecolor=black,        % color of links to bibliography
filecolor=black,      % color of file links
urlcolor=black,           % color of external links
runcolor=black
}

\usepackage[cmex10]{amsmath}
\usepackage{bbm}
\usepackage{algorithm}
\usepackage{algorithmic}
\usepackage{array}
\usepackage{url}
\begin{document}

%\title{Trade-offs between AC and reserve feasibility in unit commitment with AC power flow}
%\title{Exploring the Trade-offs between AC and reserve feasibility in unit commitment with AC power flow}
\title{Managing Power Balance and Reserve Feasibility in the AC Unit Commitment Problem}

%% To specify the authors when (number of affiliations <= 2)
\author{
\IEEEauthorblockN{Robert Parker and Carleton Coffrin}
\IEEEauthorblockA{Los Alamos National Laboratory \\
  Los Alamos, New Mexico, USA\\
\{rbparker,cjc\}@lanl.gov}
}

\maketitle

\begin{abstract}
%The Latex template and basic guidelines for the preparation of a technical paper 
%for the PSCC 2024 conference are presented. The abstract is limited to 150 words
%and cannot contain equations, figures, tables, or references. It should concisely
%state what was done, how it was done, principal results, and their significance.
Incorporating the AC power flow equations into unit commitment models
has the potential to avoid costly corrective actions required by less accurate power flow
approximations. However, research on unit commitment with AC power flow constraints
has been limited to a few relatively small test networks.
This work investigates large-scale AC unit commitment problems for the day-ahead market
and develops decomposition algorithms capable of obtaining high-quality solutions at
industry-relevant scales. The results illustrate that a simple algorithm that only
seeks to satisfy unit commitment, reserve, and AC power balance constraints can obtain
surprisingly high-quality solutions to this AC unit commitment problem.
However, a naive strategy that prioritizes reserve feasibility leads to AC infeasibility,
motivating the need to design heuristics that can effectively balance reserve and AC feasibility.
Finally, this work explores a parallel decomposition strategy that allows the proposed algorithm to obtain feasible solutions on large cases within the two hour time limit required by typical day-ahead market operations.
%A modification of the algorithm reveals that these goals can conflict if not considered
%simultaneously, and a heuristic to balance reserve and AC feasibility is introduced.
%This work demonstrates a potential bottleneck that developers seeking to solve the
%AC unit commitment problem must address and provides a heuristic strategy for
%circumventing this pitfall.
\end{abstract}


% Use this to place sponsorships
%\thanksto{\noindent Submitted to the 23rd Power Systems Computation Conference (PSCC 2024).}

\section*{Nomenclature}
\subsection*{Sets}
\begin{description}
  \item[$T$] Set of time periods $t$
  \item[$J$] Set of dispatchable devices $j$
  \item[$J^{\rm \{cs,pr\}}$] ~~Set of \{consuming, producing\} devices
  \item[$J^{\rm \{cs,pr\}}_i$] ~~Set of \{consuming, producing\} devices at bus $i$
  %\item[$J^{\rm pr}$] Set of producing devices
  %\item[$J^{\rm pr}_i$] Set of producing devices at bus $i$
  %\item[$J^{\rm sh}_i$] Set of shunt devices at bus $i$
  \item[$J^{\rm ac}$] Set of AC transmission lines
  \item[$I$] Set of buses $i$
  \item[$J^{\rm \{fr,to\}}_i$] ~~Set of AC transmission lines \{from, to\} bus $i$
  %\item[$J^{\rm to}_i$] Set of AC transmission lines to bus $i$
  \item[$N^p$] Set of active reserve zones $n$
  \item[$J_n^{\rm pr,cs}$] Set of producing {\bf and} consuming devices in
    active reserve zone $n$
\end{description}

\subsection*{Parameters}
\begin{description}
  \item[$d_t$] Duration of time period $t$
  \item[$p^{\rm min}_{jt}$] Minimum online power of device $j$ at $t$
  \item[$p^{\rm max}_{jt}$] Maximum online power of device $j$ at $t$
  \item[$p^{\rm ru}_{jt}$] Maximum online ramp-up rate of device $j$ at $t$
  \item[$p^{\rm ru,su}_{jt}$] Maximum start-up ramp rate of device $j$ at $t$
  \item[$p^{\rm rd}_{jt}$] Maximum online ramp-down rate of device $j$ at $t$
  \item[$p^{\rm rd,sd}_{jt}$] Maximum shut-down ramp rate of device $j$ at $t$
  \item[$g_j^{\rm sr}$] Series conductance of line $j$
  \item[$g_j^{\rm \{fr,to\}}$] ~``\{From, To\}''-side conductance of line $j$
  %\item[$g_j^{\rm to}$] ``To''-side conductance of line $j$
  \item[$b_j^{\rm sr}$] Series susceptance of line $j$
  \item[$b_j^{\rm \{fr,to\}}$] ~``\{From, To\}''-side susceptance of line $j$
  %\item[$b_j^{\rm to}$] ``To''-side susceptance of line $j$
  \item[$b_j^{\rm ch}$] Charging susceptance of line $j$
  \item[$p^{\rm rgu,req}_{nt}$] ~Regulation up reserve requirement in
    zone $n$ at $t$
  \item[$p^{\rm scr,req}_{nt}$] ~Synchronous reserve requirement in
    zone $n$ at $t$
\end{description}

\subsection*{Binary variables}
\begin{description}
  \item[$u^{\rm on}_{jt}$] On status of device $j$ at $t$
  \item[$u^{\rm \{su,sd\}}_{jt}$] ~\{Start-up, Shut-down\} status of device $j$ at $t$
  %\item[$u^{\rm sd}_{jt}$] Shut-down status of device $j$ at $t$
\end{description}

\subsection*{Continuous variables}
\begin{description}
  \item[$p_{jt}$] Dispatched real power of device $j$ at $t$
  \item[$p_{it}$] Real power mismatch at bus $i$ at $t$
  \item[$q_{it}$] Reactive power mismatch at bus $i$ at $t$
  \item[$p_{jt}^{\rm \{fr,to\}}$] ~Real power through the ``\{from, to\}'' side of line $j$
  %\item[$p_{jt}^{\rm to}$] Real power through the ``to'' side of line $j$
  \item[$q_{jt}^{\rm \{fr,to\}}$] ~Reactive power through the ``\{from, to\}'' side of line $j$
  %\item[$q_{jt}^{\rm to}$] Reactive power through the ``to'' side of line $j$
  \item[$z_{jt}$] Operating cost of device $j$ at $t$
  \item[$v_{it}$] Voltage magnitude of bus $i$ at $t$
  \item[$\theta_{it}$] Voltage angle of bus $i$ at $t$
  %\item[$\phi_{jt}$] Phase difference of transformer $j$ at time $t$
  \item[$p^{\rm rgu}_{jt}$] Regulation up reserve provided by device $j$ at $t$
  \item[$p^{\rm scr}_{jt}$] Synchronous reserve provided by device $j$ at $t$
  \item[$p^{\rm rru,on}_{jt}$] Online ramp-up reserve provided by device $j$ at $t$
  \item[$p^{\rm rgu,+}_{nt}$] Regulation up reserve shortfall in zone $n$ at $t$
  \item[$p^{\rm scr,+}_{nt}$] Synchronous reserve shortfall in zone $n$ at $t$
\end{description}

\section{Introduction}

The unit commitment problem is a power system scheduling task that is solved in day-ahead
and real-time power markets to balance supply with demand at low cost and ensure feasibility
with respect to physical, reliability, and operational constraints \cite{knueven2019}.
Due to the size of networks that grid operators consider (thousands
to tens of thousands of buses), discrete commitment decisions,
nonlinear alternating current (AC) physics that must be respected,
and strict time limits within which commitment decisions must be made,
solving the unit commitment problem is a challenging optimization task.
The current standard practice is for market operators
to use a direct current (DC) approximation of the power flow physics, which allows
the unit commitment problem to be formulated as a mixed-integer linear program
(MIP) and solved with established and performant commercial solvers such as CPLEX, Gurobi,
and Xpress \cite{oniell2011recent}.
% TODO: Potentially discuss MIP decomposition methods here, but I don't think this
% adds much to the paper. -RP
%{\color{blue}An additional advantage of MIP-based solution methods is the availability of efficient
%decomposition methods, such as Benders and Lagrangian decompositions, for handling
%large-scale security constrained problems \cite{fu2013}.}

While MIP technology has proven effective in saving hundreds of millions of
dollars annually in U.S. markets compared to previously used less accurate Lagrangian relaxation approaches \cite{oniell2011recent,carlson2012miso},
MIP-based unit commitment with a DC approximation
requires post-market clearing corrective action to ensure AC feasibility.
In a study on the 24-bus IEEE RTS test network, Castillo et al. estimate a 1\% cost
increase due to these corrections compared to the solution of
the unit commitment problem with a full AC power flow model \cite{castillo2016acuc}.
% Removed for space. -RP
%While solving the unit commitment problem with AC power flow constraints (AC
%Unit Commitment, or AC-UC) can lead to more efficient power grid operation,
However, unit commitment with AC power flow constraints (AC Unit Commitment, or AC-UC)
on industry-scale networks with practical time limits have not been widely demonstrated.

To investigate whether AC-UC problems can be solved for large-scale
networks within realistic time limits, the U.S. Department of Energy's Advanced Research
Projects Agency-Energy (ARPA-E) has developed the Grid Optimization Competition (GOC),
in which teams compete to develop solvers for the AC-UC problem as specified in the GOC Challenge 3 problem formulation \cite{goc3formulation}.
This problem formulation strives to be representative of modern power market optimization models and includes unit commitment and ramping constraints, AC power flow, variable loads, real and reactive reserve products, startup/shutdown trajectories, line switching, and contingency constraints.
The full optimization problem as written in \cite{goc3formulation} is a large-scale nonconvex, multi-period, multi-scenario mixed-integer nonlinear program (MINLP).
%Solving this problem, locally or globally, is well beyond the scope of off-the-shelf optimization solvers.
To realize the benefits of using this AC-UC problem in real-world electricity markets,
high-quality solutions to large-scale instances of this problem must be computed within realistic time limits.
% Removed for space. -RP
%The ultimate of goal of the ARPA-E's Grid Optimization Competition was to investigate if the design of such algorithms was possible.

Using the Grid Optimization Competition challenge 3 problem as foundation, this work aims to develop the simplest algorithm capable of obtaining high-quality
solutions to the AC-UC problem for the day-ahead market within the two hour time limit prescribed by the competition rules.
This ``benchmark algorithm'' is intended to help inform researchers and market software providers about what features of the AC-UC problem are necessary for a viable solution approach.
To this end, the results presented in this paper indicate that:
(1) off-the shelf optimization solvers are incapable of solving the full GOC challenge 3 problem within specified time limits;
(2) it is possible with current optimization methods (e.g., Gurobi and Ipopt) to develop high-quality heuristics for the AC-UC problem, which can solve industry-scale instances within the reasonable time limits;
(3) one of the key challenges in developing such heuristics is to manage the competing requirements of the AC power balance and reserve allocation constraints;
(4) problem decomposition and parallelization across multiple cores is essential to achieving the runtime requirements in large datasets with thousands of buses.

The next section begins by introducing the core features of the AC-UC problem that is considered in this work and motivates the aspects of the problem that make it challenging in practice.
% Removed for space. -RP
%That is followed by developing decomposition methods and heuristic algorithms to make the problem more tractable in Section \ref{sec:methods}.
%The proposed approaches are analyzed on 26 cases ranging from 70 to 8000 buses in Section \ref{sec:results} with final remarks in Section \ref{sec:conclusion} concluding the paper.

\section{Problem formulation and scale}
\label{sec:prob}

% The day-ahead unit commitment formulation from GOC Competition 3 seeks to maximize
% a market surplus objective by scheduling dispatchable devices (i.e., generators and
% variable loads) over a 48 hour time horizon defined on a 1 hour time resolution
% (i.e. $|T| = 48$).
% Traditional features of a unit commitment problem include semi-continuous power
% generation or consumption variables
% (\ref{eqn:semicontinuous}), ramping constraints (\ref{eqn:ramp-up}) and (\ref{eqn:ramp-down}),
% and piecewise-linear cost functions.

The unit commitment formulation considered maximizes market surplus
by scheduling dispatchable devices (generators and dispatchable loads)
over a 48 hour time horizon (i.e. $T=\{1,\dots,48\}$).
%Traditional features of a unit commitment
%problem include semi-continuous power generation or consumption variables
%(\ref{eqn:semicontinuous}),
%ramping constraints (\ref{eqn:ramp-up}) and (\ref{eqn:ramp-down}),
%and piecewise-linear cost functions.
%Additionally, unit commitment formulations incorporate constraints to ensure power
%supply matches power demand over (an approximation of) the network. The network
%model used can be copper-plate, a DC approximation, or an AC power flow model.
A simplified version of this unit commitment problem is given by Equation (\ref{eqn:simple-uc}).
%A simple unit commitment problem formulation with a copper-plate power balance
%is given in (\ref{eqn:simple-uc}), where $f_z$ is a convex piecewise linear function.
\begin{equation}
  \begin{array}{cll}
    \underset{p_{jt}, u_{jt}}{\max} & \displaystyle\sum_{t\in T} d_t \left(\sum_{j\in J^{\rm cs}} z_{jt} - \sum_{j\in J^{\rm pr}} z_{jt} \right) \\
    \text{s.t.} & \displaystyle\sum_{j\in J^{\rm cs}} p_{jt} = \displaystyle\sum_{j\in J^{\rm pr}} p_{jt} & \forall t\in T \\
  & \text{Constraints } (\ref{eqn:semicontinuous})-(\ref{eqn:ramp-down}) \\
  & z_{jt} = f_{z_j}( p_{jt}) & \forall t\in T, j\in J \\
  & u_{jt}^\text{on}, u_{jt}^\text{su}, u_{jt}^\text{sd} \in \{0, 1\}, p_{jt}\in \mathbbm{R} & \forall t\in T, j\in J \\
\end{array}
  \label{eqn:simple-uc}
\end{equation}
This formulation contains a copper-plate real power balance, 
semi-continuous real power constraints (\ref{eqn:semicontinuous}),
ramping constraints (\ref{eqn:ramp-up}) and (\ref{eqn:ramp-down}),
and a convex piecewise-linear function $f_{z_j}$ for the operating cost of each device.
Here, binary variables $u^{\rm on}_{jt}$, $u^{\rm su}_{jt}$, and $u^{\rm sd}_{jt}$
encode a device's status as either ``on'', ``starting up'', or ``shutting down''.
For example, $u^{\rm on}_{jt}=1$ means that device $j$ is scheduled to be online at time
period $t$. In this case, the device's real power $p_{jt}$ must be within its bounds
$p^\text{min}_{jt}$ and $p^\text{max}_{jt}$. If $u^{\rm on}_{jt}=0$, the device's real
power must be zero.
Binary variables for start-up and shut-down are used to encode different maximum
ramp rates for devices in these states, as implemented in (\ref{eqn:ramp-up}) and
(\ref{eqn:ramp-down}).
For more information on unit commitment problems and formulations, see
\cite{knueven2019,anjos2017,carrion2006,meibom2011advanced}.
% Removed for space. -RP
%For information on the typical inputs and outputs of a unit commitment problem, see \cite{meibom2011advanced}.
\begin{equation}
  u^\text{on}_{jt} p_{jt}^\text{min} \leq p_{jt} \leq u^\text{on}_{jt} p_{jt}^\text{max} ~\forall t\in T, j\in J \\
  \label{eqn:semicontinuous}
\end{equation}
\begin{equation}
  p_{jt} - p_{j,t-1} \leq
  \begin{array}{l}
    d_t \left(p^\text{ru}_j\left(u^\text{on}_{jt} - u^\text{su}_{jt}\right)\right.\\
    ~ ~\left.+ p^\text{ru,su}_j\left(u^\text{su}_{jt} + 1 - u^\text{on}_{jt}\right)\right)\\
  \end{array} ~\forall t\in T, j\in J
  \label{eqn:ramp-up}
\end{equation}
\begin{equation}
  p_{jt} - p_{j,t-1} \geq
    -d_t \left(p^\text{rd}_j u^\text{on}_{jt}
    + p^\text{rd,sd}_j\left(1 - u^\text{on}_{jt}\right)\right)~\forall t\in T, j\in J
  \label{eqn:ramp-down}
\end{equation}

Binary variables $u^{\rm on}_{jt}$, $u^{\rm su}_{jt}$, and $u^{\rm sd}_{jt}$
are the primary source of discrete variables in the GOC AC-UC problem,
and ramping constraints (\ref{eqn:ramp-up}) and (\ref{eqn:ramp-down}) are the
primary source of intertemporal linking.
The full formulation of this problem can be found in the ARPA-E Grid Optimization
Competition Challenge 3 Problem Formulation \cite{goc3formulation}.
Including all specifications on input data,
this formulation contains over 300 equations and is omitted here for brevity.
The notation in this work is consistent with the problem formulation, which
may be consulted for further reference.
The remainder of this section summarizes key features of the problem
formulation that differentiate it from previous work in the area.

\subsection{AC network constraints}

The unit commitment problem in this work considers AC transmission lines
and the associated power flow (derived from Ohm's law) and power balance
(derived from Kirchhoff's first law) using the polar power-voltage
formulation.
%The power flow and balance equations are defined over the network of
%buses $I$ and AC transmission lines $J^{\rm ac}$, with $i_j^{\rm fr}$
%and $i_j^{\rm to}$ denoting the buses on either side of line $j$.
The power flow equations are defined with $i_j^{\rm fr}$ and $i_j^{\rm to}$
denoting the buses on either side of line $j$.
Real and reactive power on each side of an AC transmission
line are given by (\ref{eqn:p_fr})-(\ref{eqn:q_to}).
For more information on this power flow formulation, see \cite{cain2012history}.
The power flow equations are nonlinear, nonconvex constraints and add significant
complexity when incorporated in the unit commitment problem.
The GOC problem formulation also considers DC lines, transformer branches, and
control thereof \cite{goc3formulation}.
%The power flow equations for these branches may be found in the formulation \cite{goc3formulation}.
The algorithms presented in this work but keep transformer operating settings fixed
but are free to adjust the power flow along a DC line.
\begin{multline}
  p^{\rm fr}_{jt} = u^{\rm on}_{jt} \left(
  \left(g_j^{\rm sr} + g^{\rm fr}_j\right) {v_{it}^2}\right.
    \\
    + \left(
  -g_j^{\rm sr}\cos\left(\theta_{it} - \theta_{i't}\right)\right.
  \left.\left. -b_j^{\rm sr}\sin\left(\theta_{it} - \theta_{i't}\right)
    \right) {v_{it}v_{i't}}
  \right)
  ,\\
  \forall t\in T, j\in J^{\rm ac}, i = i_j^{\rm fr}, i' = i_j^{\rm to}
  \label{eqn:p_fr}
\end{multline}
\vspace{-0.75cm}
\begin{multline}
  q^{\rm fr}_{jt} = u^{\rm on}_{jt} \left(
    \left(-b_j^{\rm sr} - b_j^{\rm fr} - b^{\rm ch}_j/2\right) {v_{it}^2}
\right.\\
  + \left(
  b_j^{\rm sr} \cos\left(\theta_{it} - \theta_{i't}\right) \right.
  \left.\left. -g_j^{\rm sr}\sin\left(\theta_{it} - \theta_{i't}\right)
    \right) {v_{it}v_{i't}}
  \right)
  ,\\
  \forall t\in T, j\in J^{\rm ac}, i = i_j^{\rm fr}, i' = i_j^{\rm to}
  \label{eqn:q_fr}
\end{multline}
\vspace{-0.75cm}
\begin{multline}
  p^{\rm to}_{jt} = u^{\rm on}_{jt} \left(
  \left(g_j^{\rm sr} + g^{\rm to}_j\right) {v_{i't}^2}\right.
    \\
    +\left(
  -g_j^{\rm sr}\cos\left(\theta_{it} - \theta_{i't}\right)\right.
  \left.\left. -b_j^{\rm sr}\sin\left(\theta_{it} - \theta_{i't}\right)
    \right) {v_{it}v_{i't}}
  \right)
  ,\\
  \forall t\in T, j\in J^{\rm ac}, i = i_j^{\rm fr}, i' = i_j^{\rm to}
  \label{eqn:p_to}
\end{multline}
\vspace{-0.75cm}
\begin{multline}
  q^{\rm to}_{jt} = u^{\rm on}_{jt} \left(
    \left(-b_j^{\rm sr} - b_j^{\rm to} - b^{\rm ch}_j/2\right) {v_{i't}^2}
  \right.\\
  + \left(
  b_j^{\rm sr} \cos\left(\theta_{it} - \theta_{i't}\right) \right.
  \left.\left. -g_j\sin\left(\theta_{it} - \theta_{i't}\right)
    \right) {v_{it}v_{i't}}
  \right)
  ,\\
  \forall t\in T, j\in J^{\rm ac}, i = i_j^{\rm fr}, i' = i_j^{\rm to}
  \label{eqn:q_to}
\end{multline}
Real and reactive power balance equations are given by (\ref{eqn:p-balance})
and (\ref{eqn:q-balance}). Bus power variables $p_{it}$ and $q_{it}$ are
mismatch slack variables, or power balance violations, that are penalized in the
objective function. In this way, the power balance equations are soft constraints
that always have a feasible solution.
%Mismatch variables $p_{it}$ and $q_{it}$ are also
%referred to as the power balance violations in this work.
\begin{multline}
  \sum_{j\in J_i^{\rm cs}} p_{jt} % + \sum_{j\in J_i^{\rm sh}} p_{jt}
  + \sum_{j\in J_i^{\rm fr}} p_{jt}^{\rm fr}
  + \sum_{j\in J_i^{\rm to}} p_{jt}^{\rm to}
  = \sum_{j\in J_i^{\rm pr}} p_{jt} + p_{it}\\
  \forall t\in T, i\in I
  \label{eqn:p-balance}
\end{multline}
\vspace{-0.75cm}
\begin{multline}
  \sum_{j\in J_i^{\rm cs}} q_{jt} % + \sum_{j\in J_i^{\rm sh}} q_{jt}
  + \sum_{j\in J_i^{\rm fr}} q_{jt}^{\rm fr}
  + \sum_{j\in J_i^{\rm to}} q_{jt}^{\rm to}
  = \sum_{j\in J_i^{\rm pr}} q_{jt} + q_{it}\\
  \forall t\in T, i\in I
  \label{eqn:q-balance}
\end{multline}
In the GOC problem formulation \cite{goc3formulation} the AC power flow constraints also include shunt devices which are controlled via a collection of discrete ``steps''.
However, shunt steps are considered fixed by all
algorithms in this work, so these terms are omitted for brevity.

\subsection{Reserve products}
To ensure a power network has sufficient capacity to balance small deviations
in supply and demand and to provide backup in case of an unexpected disturbance or
shut-down, dispatchable devices provide reserve products in addition to actual (i.e.,
planned) generation or consumption \cite{vandenbergh2020,bublitz2019}.
%A unit commitment problem with reserve constraints is given by \cite{morales2014}.
The unit commitment formulation considered in this work 
contains reserve products for both real
and reactive power capacity of all dispatchable devices, which includes both generators and loads.
Reserve products are categorized as ``up'' or ``down'' according to whether they are provided by a device's capacity to add or remove net power from the system.
Reserves must be satisfied within active and reactive reserve zones, each of which partition the dispatchable devices based on which network buses they are connected to.
The reserve requirements for each zone are implemented as soft constraints with a slack variable that is penalized
in the objective if the requirement is not satisfied.
The reserve balance equations for regulation up reserve and synchronous reserve are
given as examples in (\ref{eqn:rgu-balance}) and (\ref{eqn:scr-balance}).
Reserves for each device are constrained by the ``headroom'' to the device's upper
or lower bounds depending on the type of reserve and the type of device.
For example, regulation up, synchronous, and online regulation ramp-up
reserves are constrained by the headroom to a producer's upper bound, as in
(\ref{eqn:producer-headroom-max}), or the headroom to a consumer's
lower bound, as in (\ref{eqn:consumer-headroom-min}).
In all, the GOC problem formulation considers eight active reserve products
and two reactive reserve products.
In this work the vectors $p^{\rm RES}$
and $q^{\rm RES}$ are used to denote an assignment of all active and reactive reserve
products for all dispatchable devices across all time periods.
%Active reserve balance equations for an active reserve zone $n$ and
%time period $t$ are given by (\ref{eqn:rgu-balance})-(\ref{eqn:rrd-balance}),
%and reactive reserve balance equations are given by (\ref{eqn:qru-balance})
%and (\ref{eqn:qrd-balance}).
Higher-quality reserve products, such as regulation up, can be used simultaneously
for lower-quality products such as synchronized reserve, as implemented in
(\ref{eqn:rgu-balance}) and (\ref{eqn:scr-balance}).
\begin{equation}
  \sum_{j\in J_n^{\rm pr,cs}} p^{\rm rgu}_{jt} + p_{nt}^{\rm rgu,+} \geq p^{\rm rgu,req}_{nt}
  ,\hspace{0.2cm}\forall t\in T, n\in N^p
  \label{eqn:rgu-balance}
\end{equation}
\vspace{-0.5cm}
%\begin{equation}
%  \sum_{j\in J_n^{p,c}} p_{jt}^{\rm rgd} + p_{nt}^{\rm rgd,+} \geq p_{nt}^{\rm rgd,req}
%  \label{eqn:rgd-balance}
%\end{equation}
\begin{multline}
  \sum_{j\in J_n^{\rm pr, cs}} \left(p_{jt}^{\rm rgu} + p_{jt}^{\rm scr}\right)
  + p^{\rm scr,+}_{jt} \geq p_{nt}^{\rm rgu,req} + p_{nt}^{\rm scr,req}
  ,\\
  \forall t\in T, n\in N^p
  \label{eqn:scr-balance}
\end{multline}
%\begin{equation}
%  \sum_{j\in J_n^{p,c}} \left(p_{jt}^{\rm rgu} + p_{jt}^{\rm scr} + p_{jt}^{\rm nsc}\right)
%  + p^{\rm nsc,+}_{jt} \geq p_{nt}^{\rm rgu,req} + p_{nt}^{\rm scr,req} + p_{nt}^{\rm nsc,req}
%  \label{eqn:nsc-balance}
%\end{equation}
%\begin{equation}
%  \sum_{j\in J_n^{p,c}} \left(p_{jt}^{\rm rru,on} + p_{jt}^{\rm rru,off}\right)
%  + p_{nt}^{\rm rru,+} \geq p_{nt}^{\rm rru,min}
%  \label{eqn:rru-balance}
%\end{equation}
%\begin{equation}
%  \sum_{j\in J_n^{p,c}} \left(p_{jt}^{\rm rrd,on} + p_{jt}^{\rm rrd,off}\right)
%  + p_{nt}^{\rm rrd,+} \geq p_{nt}^{\rm rrd,min}
%  \label{eqn:rrd-balance}
%\end{equation}
%\begin{equation}
%  \sum_{j\in J_n^{p,c}} q_{jt}^{\rm qru} + q_{nt}^{\rm qru,+} \geq q_{nt}^{\rm qru,min}
%  \label{eqn:qru-balance}
%\end{equation}
%\begin{equation}
%  \sum_{j\in J_n^{p,c}} q_{jt}^{\rm qrd} + q_{nt}^{\rm qrd,+} \geq q_{nt}^{\rm qrd,min}
%  \label{eqn:qrd-balance}
%\end{equation}
\begin{equation}
  p + p^{\rm rgu}_{jt} + p^{\rm scr}_{jt} + p^{\rm rru,on}_{jt}
  \leq p^{\rm max}_{jt} u^{\rm on}_{jt}, \hspace{0.2cm}\forall t\in T, j\in J^{\rm pr}
  \label{eqn:producer-headroom-max}
\end{equation}
\begin{equation}
  p - p^{\rm rgu}_{jt} - p^{\rm scr}_{jt} - p^{\rm rru,on}_{jt}
  \geq p^{\rm min}_{jt} u^{\rm on}_{jt}, \hspace{0.2cm}\forall t\in T, j\in J^{\rm cs}
  \label{eqn:consumer-headroom-min}
\end{equation}

In addition to penalties incurred if reserve requirements are not satisfied, devices
may incur additional cost by providing reserves.
As reserve constraints are linear and reserve variables are continuous, the problem
of balancing reserve shortfall penalties with device reserve costs may be modeled
as a linear program (LP) for fixed power levels and commitment decisions $p$, $q$, and $u$.
While these constraints are not complicated individually, the sheer number of constraints
and variables required to encode the full reserve model
(five headroom constraints and ten reserve variables per device)
adds significant complexity to the AC-UC problem.

%\subsection{Dispatchable loads}
%
%In contrast to many common optimal power flow formulations, the Grid
%Optimization Competition considers commitment of both dispatchable
%generators and dispatchable loads.
%

\subsection{Contingency constraints}

Security constraints are an important feature of electric power markets, and may be
incorporated into Unit Commitment problems \cite{wu2007} or Optimal Power Flow
problems \cite{aravena2023}.
The GOC AC-UC problem considers security constraints that penalize branch flow
thermal limit violations in a defined set of contingencies, where each contingency
is defined as the loss of a single specified branch in the network.
While solutions computed in this work are evaluated in the context of these
contingency constraints, the algorithms presented do not account for them
as they did not contribute to significant penalties in the objective function
value.

%\rev{I am thinking we need to mention this here; as it does appear in the competition
%formulation. However, it may be reasonable to only discuss with text and not equations
%and mention that this feature was not explicitly modeled due to not yielding as notable
%violations in the solver.}

\subsection{Problem scale}

%AC Unit Commitment problems have been studied in previous literature.
Early approaches to AC Unit Commitment applied Lagrangian relaxation \cite{sanchez1999}
and Benders decomposition \cite{ma1999} to problems defined on the
IEEE 118-bus network. Fu et al. have considered a security-constrained version of this
problem on the same network \cite{fu2005} using an augmented
Lagrangian and Benders decomposition approach.
Castillo et al. apply an outer approximation method to solve AC-UC
problems on networks of up to 118 buses and compare with a method that employs
DC unit commitment followed by an AC feasibility solve \cite{castillo2016acuc},
while Tejada-Arango et al. benchmark a direct mixed integer nonlinear programming
(MINLP) formulation, an approach based on sequential linear programming, and
a second-order conic programming formulation \cite{arango2019}.
These previous works on the AC-UC problem have applied exact local MINLP algorithms
to this problem and have only demonstrated scalability up to power networks of on the
order of 100 buses and 24 time points, which is far from the size desired by industrial practitioners.
Section \ref{sec:solver-eval} investigates the scalability of the Knitro \cite{knitro} MINLP
solver on AC-UC problems and confirms that large-scale instances are out of reach.
%Focusing on a practical deployment scenario,
This work considers AC-UC problems on the Grid Optimization Competition ranging from
73 buses to more than 8,000 buses with 48 time points,
and approaches solving such large problems by decomposing the full AC-UC problem into
several subroutines and parallelizing these tasks across multiple cores. % which are discussed in the next section.

\section{Methods}
\label{sec:methods}

The decomposition algorithms that are developed in this work rely on several subroutines,
each focusing on different aspects of the complete AC-UC problem.
%, such as temporal scheduling, AC optimal power flow, and reserve allocation.
This section introduces these subroutines then describes how they are combined into four distinct heuristic solution approaches.

\subsection{Subroutines}
\label{sec:subroutines}

\subsubsection{Copper-plate scheduling}
\label{sec:copper-plate-uc}
The unit commitment approach taken is to model the full-horizon scheduling problem
as a MIP. For tractability, the scheduling problem
considers only copper-plate real and reactive power balances.
That is, no network model is considered. Only total demand and generation
  must match, as if all devices are connected to a single conductive plate.
The full zonal reserve requirement model is considered in this subproblem as well.
The outputs of this subproblem are discrete commitment decisions $u$, real and reactive
power estimates $p$ and $q$, and real and reactive reserve commitments
$p^{\rm RES}$ and $q^{\rm RES}$.

\subsubsection{AC optimal power flow (AC-OPF)}
This work achieves AC feasibility by solving an AC optimal power flow (AC-OPF)
problem at individual time periods $t$. Each subproblem has the discrete commitment decisions fixed and only considers devices that are online in time period $t$.
Device costs are convex piecewise-linear cost functions modeled with the
$\Delta$ formulation described by \cite{coffrin2021piecewise}.
In this work, shunts and transformer taps are fixed to their operating
conditions at the initial point.
To keep AC-OPF problem sizes manageable, reserve constraints are not considered
by these subproblems. Incorporating these constraints would add ten additional
variables and five additional inequality constraints per dispatchable device
and several additional variables and inequality constraints per reserve zone, nearly doubling the size of the model.

\subsubsection{Reserve allocation}
\label{sec:reserve-allocation}
While reserve products and zonal reserve requirements are incorporated in the
unit commitment model, they are not explicitly considered by the AC-OPF model.
For this reason, care is taken to reallocate feasible reserves after the commitment schedule and AC-feasible power dispatch is generated.
The algorithms discussed in Section \ref{sec:algorithms} differ primarily in how they allocate reserve products based on the methods discussed here.

{\em Greedy reserve allocation,} the first method considered for allocating reserves, is a simple greedy procedure that, for every dispatchable device, attempts to allocate the headroom to
the device's upper and lower bounds to the appropriate reserve products, starting with the highest value products.
In this approach, the cost of providing a reserve and the zonal reserve requirements are not considered.
That is, reserve is assigned for every device possible even if doing so is not necessary to satisfy the requirements.
This is motivated by the observation that, in most cases, the penalty for failing to meet a reserve requirement is much larger than the cost of providing the reserve, which is often zero.

{\em Tighten device bounds,} the second method considered, fixes real reserve products computed by the
copper-plate unit commitment schedule on all devices where doing so does not
lead to a guaranteed local power balance violation on the device's bus.
A local power balance violation occurs if, for example, a lone producing device
on a bus is scheduled to provide more regulation down reserve than the sum
of upper bounds on real power that can be transmitted from adjacent lines.
For devices where this type of violation does not occur, bounds are tightened
on device real power variables $p$ to guarantee that the reserves committed
by the unit commitment problem are feasible after the AC-OPF subproblems.
If a large number of devices provide reserves, this method can become overly
restrictive for the AC-OPF subproblems and can lead to power balance violations.
For this reason, a variation of this method is considered where only some
fraction $\gamma$ of devices providing real power reserve have their bounds
tightened. When $\gamma < 1$, devices are sorted by the value of reserve that
they provide and the top $\gamma$ of them are constrained in the AC-OPF subproblems.
This method can be also be used in conjunction with post-AC-OPF reserve allocation.

{\em Reserve re-dispatch via linear programs,} the final method, computes reserve products post-AC-OPF by fixing commitment
decisions $u$ and dispatched real and reactive power levels $p$ and $q$ by
solving the reserve model (zonal reserve balances, headroom constraints,
costs, and penalties) as a linear program (LP). These LPs are independent
at each time period and are solved in parallel.

\subsection{Decomposition algorithms}
\label{sec:algorithms}
The subroutines described in Section \ref{sec:subroutines} are combined into the follow four algorithms for solving the full AC-UC problem.

Algorithm \ref{alg:reserve-preserving} is a simple decomposition designed
to mimic a basic version of current industrial practice. % Cite FERC 2011 report?
A commitment schedule $u$, along with nominal real power $p$, reactive power $q$,
and real reserve products $p^{\rm RES}$, is computed via the copper-plate unit
commitment problem described in Section \ref{sec:copper-plate-uc}.
Co-optimization of commitment decisions with real power reserves is chosen to mimic
a joint reserve scheduling market \cite{gonzalez2014}.
Real reserve products are then fixed, imposing tightened bounds on real power.
AC-OPF subproblems are solved sequentially, with a bound-tightening step before
each solve to ensure feasibility of ramping constraints.
Reactive power reserve products are allocated after solving AC-OPF subproblems
by the greedy strategy described in Section \ref{sec:reserve-allocation}.
\begin{algorithm}[ht!]
  \begin{algorithmic}[1]
    \STATE {\bf Initialize:} $p,q,u\gets$ initial status
    \STATE $p, q, u, p^{\rm RES}\gets$ Copper-plate unit commitment
    \STATE Tighten bounds on $p$ by fixing $p^{\rm RES}$
    \FOR{$t = 1\dots48$}
      \STATE Tighten bounds on $p_t$ using ramping constraints
      \STATE $p_t, q_t \leftarrow$ AC-OPF at $t$
    \ENDFOR
    \STATE $q^{\rm RES} \gets$ Greedy reserve allocation
    \RETURN $p, q, u, p^{\rm RES}, q^{\rm RES}$
  \end{algorithmic}
  \caption{Reserve-preserving decomposition}
  \label{alg:reserve-preserving}
\end{algorithm}

Algorithm \ref{alg:sequential-greedy} is designed to prioritize unit commitment-feasible,
AC-feasible solutions. It solves a copper-plate unit commitment problem to obtain a
commitment schedule and nominal real and reactive powers, then solves AC-OPF subproblems
sequentially. After solving AC-OPF subproblems, real and reactive powers are fixed and
both real and reactive reserve products are allocated using the greedy approach.
\begin{algorithm}[ht!]
  \begin{algorithmic}[1]
    \STATE {\bf Initialize:} $p,q,u\gets$ initial status
    \STATE $p, q, u\gets$ Copper-plate unit commitment
    \FOR{$t = 1\dots48$}
      \STATE Tighten bounds on $p_t$ using ramping constraints
      \STATE $p_t, q_t \leftarrow$ AC-OPF at $t$
    \ENDFOR
    \STATE $p^{\rm RES}, q^{\rm RES} \gets$ Greedy reserve allocation
    \RETURN $p, q, u, p^{\rm RES}, q^{\rm RES}$
  \end{algorithmic}
  \caption{Simple greedy decomposition}
  \label{alg:sequential-greedy}
\end{algorithm}

Algorithm \ref{alg:balancing} uses the same basic decomposition, but is refined
by the observation that tightening bounds on a large number of devices can
lead to AC infeasibility. It employs the same bound tightening strategy as
Algorithm \ref{alg:reserve-preserving} but only for the top 5\% of devices
providing reserve value, as described in Section \ref{sec:reserve-allocation}.
After AC-OPF subproblems, reserves are re-computed via linear programs that
balance costs with zonal reserve penalties subject to fixed real power, reactive
power, and on status variables previously computed.
\begin{algorithm}[h!]
  \begin{algorithmic}[1]
    \STATE {\bf Parameter} $\gamma = 5$
    \STATE {\bf Initialize:} $p,q,u\gets$ initial status
    \STATE $p, q, u, p^{\rm RES}, q^{\rm RES} \gets$ Copper-plate unit commitment
    \STATE Tighten $p$ bounds by fixing $p^{\rm RES}$ for top $\gamma\%$ of devices
    \FOR{$t = 1\dots48$}
      \STATE Tighten bounds on $p_t$ using ramping constraints
      \STATE $p_t, q_t \leftarrow$ AC-OPF at $t$
    \ENDFOR
    \STATE $p^{\rm RES}, q^{\rm RES} \gets$ Reserve re-dispatch via linear programs
    \RETURN $p, q, u, p^{\rm RES}, q^{\rm RES}$
  \end{algorithmic}
  \caption{Reserve/AC-balancing heuristic}
  \label{alg:balancing}
\end{algorithm}

Algorithm \ref{alg:parallel} is the balancing heuristic algorithm with a parallel
decomposition of the AC-OPF subproblems at individual time periods.
This is followed by a sequential projection of real power into the bounds implied by
ramping constraints at the previous time point. The resulting solutions satisfy
the unit commitment and ramping constraints, have modest AC power balance
violations due to ramping constraint projections, and have reserves that are
balanced between the pre-OPF copper-plate reserve schedule and the post-OPF reserve re-dispatch.
This algorithm variant is comparable to the solution method that was used for the
``ARPA-E Benchmark'' algorithm in Event 4 of GOC Challenge 3.
\begin{algorithm}[h!]
  \begin{algorithmic}[1]
    \STATE {\bf Parameter} $\gamma = 5$
    \STATE {\bf Initialize:} $p,q,u\gets$ initial status
    \STATE $p, q, u, p^{\rm RES}, q^{\rm RES} \gets$ Copper-plate unit commitment
    \STATE Tighten $p$ bounds by fixing $p^{\rm RES}$ for top $\gamma\%$ of devices
    \FOR{$t = 1\dots48$ in parallel}
      \STATE $p_t, q_t \leftarrow$ AC-OPF at $t$
    \ENDFOR
    \FOR{$t = 1\dots48$}
      \STATE Project $p_t$ to satisfy ramping constraints
    \ENDFOR
    \STATE $p^{\rm RES}, q^{\rm RES} \gets$ Reserve re-dispatch via linear programs
    \RETURN $p, q, u, p^{\rm RES}, q^{\rm RES}$
  \end{algorithmic}
  \caption{Parallel decomposition heuristic}
  \label{alg:parallel}
\end{algorithm}

These decomposition algorithms consider all hard constraints of the GOC problem
formulation \cite{goc3formulation}. That is, omitted features such as contingency constraints
are modeled as soft constraints. As such, solutions produced by these algorithms are
guaranteed to be feasible for the GOC problem formulation.
However, as all of these algorithms are heuristics, they provide no particular guarantees
on solution speed or quality. 
To address this, future work may focus on using these decompositions in
  a Benders decomposition or outer approximation framework to develop approaches
that, given enough time, are guaranteed to converge to a locally optimal solution.
The next section conducts detailed experiments on a variety of realistic network instances
to explore the strengths and weaknesses of these algorithms.  
%\rev{prep for next section}.

\section{Results and discussion}
\label{sec:results}

\subsection{Full AC-UC problem evaluation}
\label{sec:solver-eval}

To motivate the use of heuristic decomposition algorithms for large scale instances
of the AC-UC problem, we first evaluate the performance of Knitro 14.0 on small instances
of the full AC-UC problem,
using a heuristic branch-and-bound algorithm \cite{queseda1992} with nonlinear programming (NLP)
subproblems solved by an interior point method \cite{waltz2006}.
We use small networks %from the ``sandbox'' dataset provided by the Grid Optimization Competition,
and construct an AC-UC problem containing only unit commitment and AC network constraints
({\it i.e.} omitting reserve products).
The test datasets have 3, 14, 37, 73, and 617 buses, and come either from GOC Event 4
or the ``sandbox'' dataset, indicated by ``S0'' in the Case ID.
Solve times and objective values for these small instances of the simplified
problem are shown in Table \ref{tab:knitro-acuc}.
The results indicate that, even for this simplified problem, Knitro is unable to
solve a 617-bus instance of the full AC-UC problem as an MINLP with 627,768 variables and 705,544 constraints.
The largest instance considered in this work, with 8,316 buses, has 8,260,224 variables and 8,947,330 constraints
in this simplified version of the AC-UC problem.
This is well beyond the scope of off-the-shelf
MINLP solvers, and motivates the use of heuristic decomposition methods to produce
high-quality solutions for industry-scale instances of this problem.

\begin{table}
  \centering
  \caption{Results of solving full AC-UC problem with Knitro}
  \begin{tabular}{ccccc}
    \hline\hline
    Case ID & N. Var. & N. Con. & Objective (\$) & Solve time (s) \\
    \hline\hline
    S0N00003-003 & 2260 & 3850 & 9.07e+05 & 71\\
    S0N00014-003 & 22080 & 26183 & 2.23e+06 & 33\\
    S0N00037-003 & 43584 & 47494 & 1.07e+07 & 20\\
    %S0N00073-002 & 157920 & 197808 & N/A & $>$10000\\
    N00073-333 & 159072 & 195415 & 2.30e+08 & 84 \\
    N00617-002 & 627768 & 705544 & N/A & $>$10000 \\
    \hline\hline
  \end{tabular}
  \label{tab:knitro-acuc}
\end{table}
% This is an alternative table where we present the solve time of only the
% root node relaxation.
%\begin{table}
%  \centering
%  \caption{Results of solving full AC-UC problem with Knitro}
%  \begin{tabular}{ccccc}
%    \hline\hline
%    Case ID & N. Var. & N. Con. & Objective (\$) & Root node time (s) \\
%    \hline\hline
%    S0N00003-003 & 2260 & 3850 & 9.07e+05 & $<$1\\
%    S0N00014-003 & 22080 & 26183 & 2.23e+06 & 1\\
%    S0N00037-003 & 43584 & 47494 & 1.07e+07 & 4\\
%    N00073-333 & 159072 & 195415 & 2.30e+08 & 20 \\
%    N00617-002 & 627768 & 705544 & N/A & 6760 \\
%    N02000-002 & 627768 & 705544 & N/A & $>$10000 \\
%    \hline\hline
%  \end{tabular}
%  \label{tab:knitro-acuc}
%\end{table}

\subsection{Test data}

The algorithms presented in Section \ref{sec:algorithms} are evaluated in terms of
solution quality and solve time on 28 day-ahead AC-UC cases provided by Event 4
of the Grid Optimization Competition, Challenge 3.
Networks with 73, 617, 2,000, 4,224, 6,049, 6,717, and 8,316 buses are considered.
Four representative scenarios are considered for each network.
Problem data for each network are given in Table \ref{tab:network-stats}.
Except for the 73-bus network, all scenarios for a given network have the same numbers
of dispatchable devices, AC lines, and reserve zones.
Each case is assigned a unique identifier (``Case ID'') using the network size and the
scenario number from GOC Event 4.
%For each network and scenario, the objective value of the solution provided by
%each of the algorithms in Section \ref{sec:algorithms} is presented along with
%best-known solution from GOC Event 4 \cite{goc3leaderboards} and the gap between
%the two solutions.
The objective values of solutions are computed by an independent evaluation program
provided by the GOC that considers the full problem formulation \cite{goc3evaluator}.

\begin{table}
  \centering
  \caption{Statistics for the Event 4 networks considered}
  \begin{tabular}{ccccc}
    \hline\hline
    Buses & Devices & AC lines & $p$ res. zones & $q$ res. zones \\
    \hline\hline
    73 (scenario 991) & 208 & 127 & 1 & 1 \\
    73 (other scenarios)& 205 & 105 & 1 & 1 \\
    617 & 499 & 723 & 10 & 10 \\
    2000 & 1894 & 2345 & 4 & 10 \\
    4224 & 2151 & 2605 & 2 & 2 \\
    6049 & 3774 & 4920 & 6 & 6 \\
    6717 & 5826 & 7173 & 9 & 12 \\
    8316 & 5585 & 7723 & 7 & 7 \\
    \hline\hline
  \end{tabular}
  \label{tab:network-stats}
\end{table}

\subsection{Computational Setting}

All results in this section are computed using HPE ProLiant XL170r servers with two Intel 2.10 GHz CPUs and 128 GB of memory using Julia v1.6 \cite{julia}.
These machine have 36 physical cores and up to 72 simultaneous threads.
Optimization models are constructed with JuMP v1.14 \cite{jump}.
The %copper plate
unit commitment MIP models are solved with Gurobi v10.0 and the
AC-OPF subproblems are solved with Ipopt v3.14 \cite{ipopt} using the MA27 linear
solver \cite{ma27} and a symbolic automatic differentiation approach \cite{symbolicad}
that is similar to the approach of Gravity \cite{gravity}.
To %mitigate the performance impact
reduce the runtime
of ill-conditioned instances that require many
interior-point iterations to converge,
%a heuristic callback is used to terminate solves early
solves are terminated via callback
if they have reached a $10^{-3}$ unscaled primal residual and a 1.0
unscaled dual residual.
The reserve assignment LPs are solved using HiGHS v1.5 \cite{highs}.

Implementations of the models and subroutines used in this work can be found
at \url{https://github.com/lanl-ansi/GOC3Benchmark.jl}.

\subsection{Solution quality}

Solution quality results for Algorithm \ref{alg:reserve-preserving} are shown
in Table \ref{tab:reserve-preserving}.
In Tables \ref{tab:reserve-preserving}-\ref{tab:parallel}, ``Objective'' is the
market surplus value of the solution produced by presented algorithm, while
``Best-known solution'' is the value of the best solution produced by any team
in the GOC, according to the leaderboards \cite{goc3leaderboards}.
Each of these best solutions was obtained within a two hour time
limit on the standardized hardware used by the competition.
``Gap'' is defined as the
%difference between these solutions divided by the best-known solution value, times 100.
percent difference with respect to the best-known solution.
``Res. penalty'' is the total penalty incurred by zonal reserve shortfalls,
while ``$p$-penalty'' and ``$q$-penalty are total real and reactive power balance violation penalties.
The penalty factor is $10^6$
US dollars per power unit, so achieving a $10^2$ power balance penalty
is equivalent to a total power balance violation magnitude of $10^{-4}$.
``Solve time'' is the time elapsed between the start of Julia code execution
and the point at which the solution file is written.

While the algorithm produces acceptable
solutions for many cases, 13 cases have gaps to the best-known solution of
over 30\%. In 12 of these 13 cases (and several others), the dominant penalties
are real and reactive power balance violation penalties.
In Table \ref{tab:reserve-preserving}, rows corresponding to solution gaps
of larger than 30\% are highlighted.
This suggests that fixing reserve products for many devices
is too restrictive to achieve AC feasibility.
%even when ignoring reserve products that would force a local line limit violation.

Algorithm \ref{alg:sequential-greedy} prioritizes AC feasibility over reserve
feasibility. It includes reserves in its copper-plate unit commitment model,
but does not fix any reserve products before the AC-OPF subproblems.
Table \ref{tab:sequential-greedy} shows solution quality results for
Algorithm \ref{alg:sequential-greedy}, which produces surprisingly high-quality
solutions for a simple greedy approach.
%considering this algorithm does not consider contingency constraints,
%line switching, shunt control, or transformer control.
Except for seven cases, these solutions are computed within the 2 hour time limit.
Of the 28 cases, 12 have a gap of less than 10\% to the best-known solution.
In 9 of the remaining 16 cases, the dominant penalty is the reserve shortfall
penalty. This suggests that post-AC-OPF greedy reserve allocation is not sufficient
to minimize reserve penalty, and that solutions could be improved by including
information from the reserves computed by the unit commitment subproblem.

\begin{figure}[ht!]
  \centering
  \vspace{-0.25cm}
  \includegraphics[width=9cm]{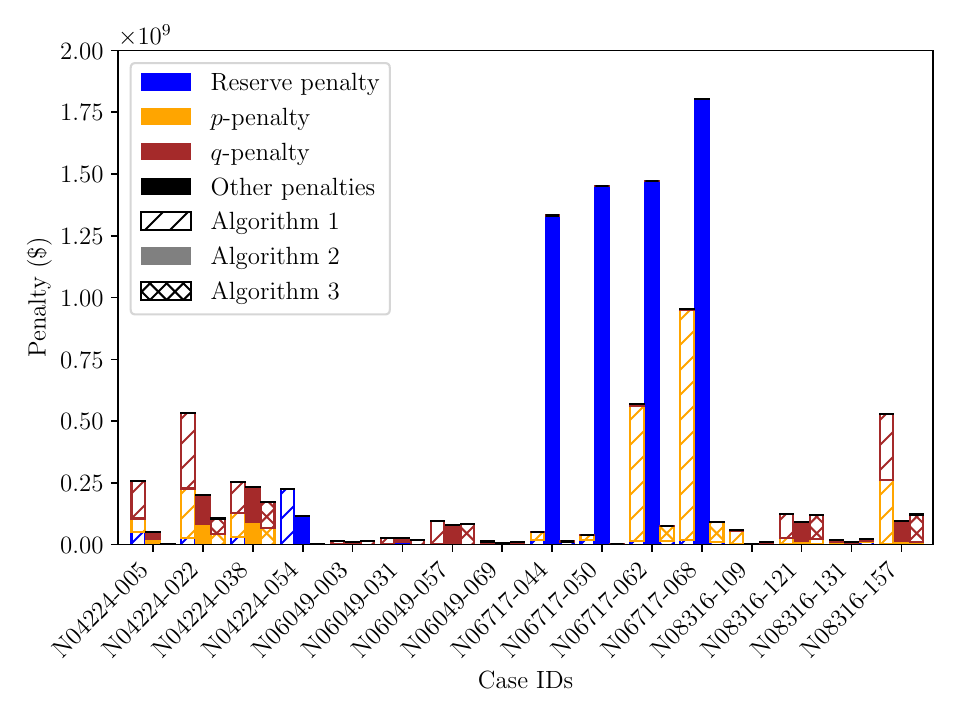}
  \vspace{-0.75cm}
  \caption{Summary of objective penalties incurred for Algorithms \ref{alg:reserve-preserving}-\ref{alg:balancing}
    on the larger networks considered in this work. Algorithm \ref{alg:balancing} reliably balances
  AC and reserve feasibility requirements, finding solutions with low penalties.}
  \vspace{-0.25cm}
  \label{fig:penalties}
\end{figure}

Algorithm \ref{alg:balancing} attempts to improve reserve shortfall penalties
by tightening bounds on $\gamma = 5\%$ of devices providing reserve value.
In addition, reserve products are re-dispatched after the AC-OPF subproblems via
linear programs with zonal balance models. The solution quality results for this algorithm
are shown in Table \ref{tab:balancing}. The results show that this algorithm produces
high-quality results in most cases, with the gap to the best-known solution less than
10\% for 19 of 28 cases. Reserve penalty dominates in only one of the remaining nine
cases (N02000-031).
The total market surplus across all cases is 1.74$\times$10$^{10}$\$
for Algorithm \ref{alg:balancing}, compared to 1.37$\times$10$^{10}$\$ for Algorithm
\ref{alg:sequential-greedy}, a 20\% improvement with a value of over 3 billion US dollars.
Fig. \ref{fig:penalties} compares the penalties incurred by Algorithms
\ref{alg:reserve-preserving}-\ref{alg:balancing} on cases 4,000 buses or larger.
The comparison shows large power balance violation penalties for Algorithm
\ref{alg:reserve-preserving} and large reserve shortfall penalties for Algorithm
\ref{alg:sequential-greedy}. In the few cases where Algorithm \ref{alg:balancing}
incurs large penalties, they are dominated by moderate power balance violations.

The results demonstrate that surprisingly good solutions can be obtained by a simple
algorithm that does not consider contingency constraints, line switching,
shunt control, or transformer control. In 16 of 28 cases considered,
Algorithm \ref{alg:balancing} obtains solutions with objectives
within 2\% of the best-known. In case N08316-131, however, the
algorithm takes longer than the 2 hour time limit to produce a solution.
%Rows corresponding to cases that exceed the time limit are highlighted in Tables
%\ref{alg:sequential-greedy} and \ref{alg:balancing}.
%Section \ref{sec:solve-time} profiles this algorithm and presents the results of
%a parallel decomposition to improve solve times.

\subsection{Solve time}
\label{sec:solve-time}

\begin{figure}[ht!]
  \centering
  \vspace{-0.25cm}
  \includegraphics[width=7cm]{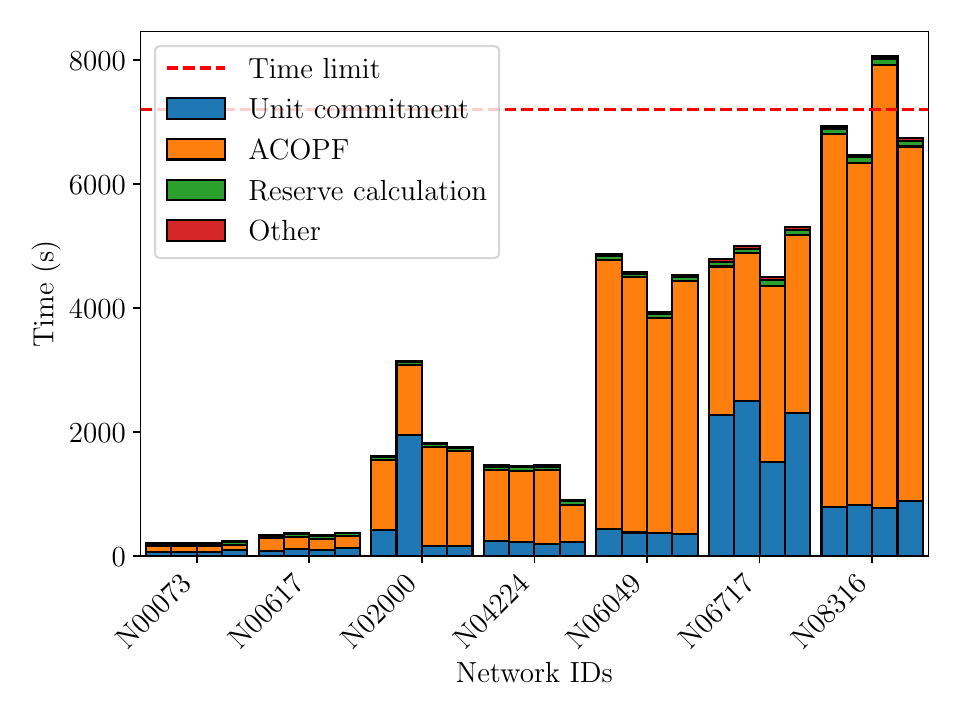}
  \vspace{-0.25cm}
  \caption{A breakdown of solve times with Algorithm \ref{alg:balancing}, where the scenarios for each network have the same order as in Table \ref{tab:balancing}. These results highlight how AC-OPF solve times are a dominant factor in the runtime of the algorithm.}
  \label{fig:time-breakdown-balancing}
\end{figure}

Fig. \ref{fig:time-breakdown-balancing} shows a breakdown of the solve times of
Algorithm \ref{alg:balancing} for the 28 cases presented.
The results show that solving the sequential AC-OPF subproblems is the dominant
computational expense, although for the 6,717-bus cases the unit commitment solve
time also contributes significantly.

Algorithm \ref{alg:parallel} solves individual AC-OPF subproblems in parallel to
attempt to reduce the dominant computational burden and provide solutions within
the two hour time limit. It does so at some expense to solution quality, however,
as the post-processing projection step that recovers ramping feasibility can also
introduce power balance violations.
Solution quality results for Algorithm \ref{alg:parallel} are shown in Table
\ref{tab:parallel}. While the parallel solution strategy does produce high-quality
solutions for most cases, it does incur a penalty compared to Algorithm
\ref{alg:balancing} on cases N04224-005, N04224-022, and N04224-038 due to power
balance violations.
% The following removed for space. -RP
%In each of these cases, relatively few projections occur.
%Case N08316-157 has the maximum number of projections with 495, out of 268,080 that
%are possible (i.e., the number of devices times the number of time periods).
%Table \ref{tab:projection} shows the number of devices projected for each of these
%cases, along with the device and time period that are most frequently projected
%for each instance.
%The projections that do occur are not evenly distributed among the dispatchable
%devices or time periods. For example, in case N06717-005, 40 projections, or 9 \%
%of the total, occur on a single device.
%That projected devices and intervals appear to be clustered suggests the presence
%of ``critical devices'' where unit commitment and AC-OPF subproblems have conflicting
%solutions and ``critical time periods'' where a large change in device power level
%is more likely to occur. Future work may attempt to predict these critical devices
%and time periods from the problem data and the unit commitment solution and develop
%strategies to mitigate their impact on solution quality.
%
% TODO: Two ``future work'' ideas for eliminating infeasibility in parallel
% decomposition:
% 1. Identify ``criticla time periods'' or ``critical devices''
% 2. Use the proposed decomposition approach in a rigorous framework like Benders,
%    Lagrangian, or ADMM

\begin{figure}[ht!]
  \centering
  \vspace{-0.25cm}
  \includegraphics[width=4.5cm]{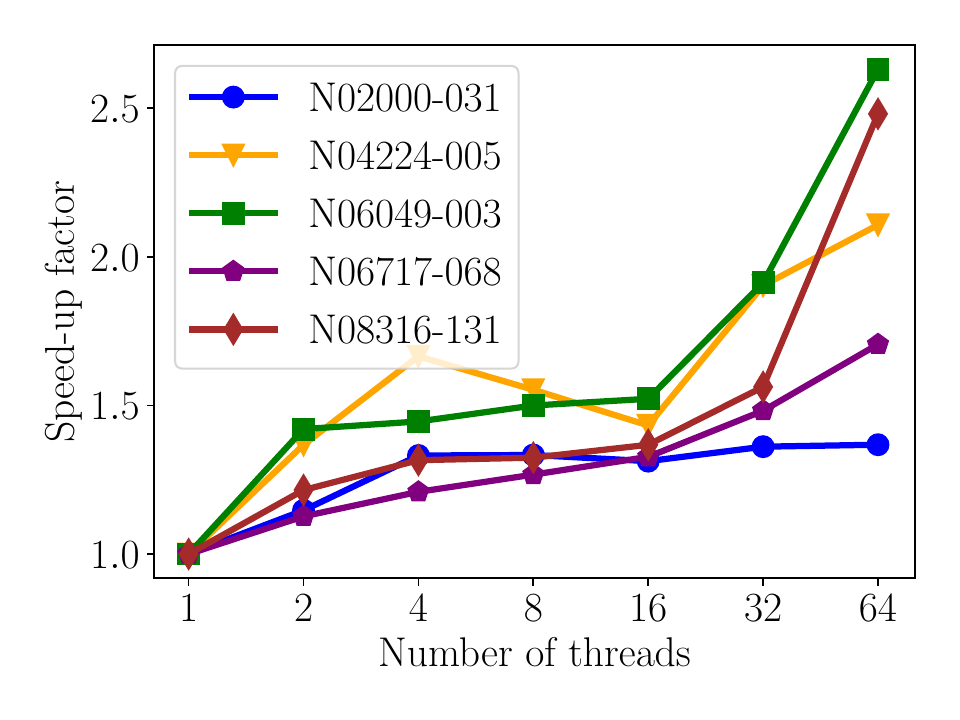}\hspace{-0.5cm}
  \includegraphics[width=4.5cm]{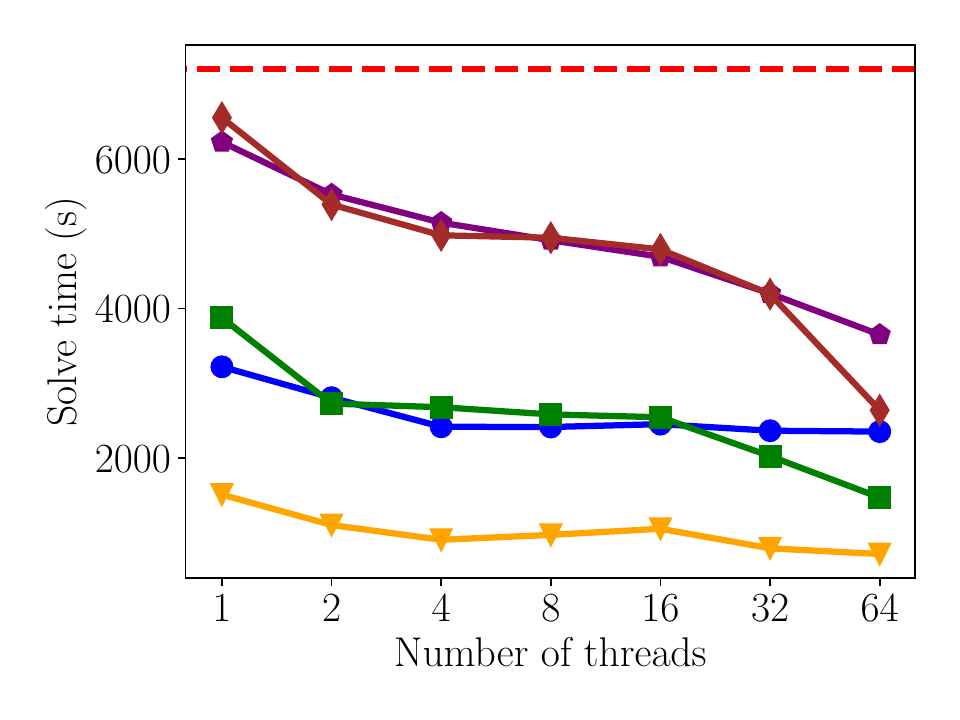}
  \vspace{-0.25cm}
  \caption{Scaling of Algorithm 4's solve time with number of threads for five of the large cases considered in this work. The results show that 32 parallel processes are sufficient for satisfying a two hour runtime limit.}
  \label{fig:parallel-scaling}
\end{figure}

Fig. \ref{fig:parallel-scaling} shows the solve times and speed-up factors as
functions of the number of threads allocated to the Julia process running
Algorithm \ref{alg:parallel} on one scenario from each of the five largest
network sizes considered.
While Algorithm \ref{alg:balancing} exceeds the time limit on case N08316-131,
Algorithm \ref{alg:parallel} solves within the time limit when using only a single
thread. With only four threads, Algorithm \ref{alg:parallel} solves within the
time limit by a margin of at least 2,000 s for all cases.
Additional runtime gains can be achieved using additional cores.
According to the results presented in Fig. \ref{fig:time-breakdown-balancing},
Algorithm \ref{alg:balancing} spends between 36\% and 90\% of its time in the sequential AC-OPF portion of the algorithm for networks 2,000 buses and larger. % with an average of 75\% 
As only this portion of the algorithm is parallelized, linear speedup with the
number of processors is not expected. A theoretical upper bound on speedup for
the most OPF-intensive instance is 10$\times$, while an upper bound on speedup for
the average instance is approximately 4$\times$.
Fig. \ref{fig:parallel-scaling} shows speedups between 1.4$\times$ and 2.6$\times$
with 64 threads for the most challenging instances, indicating the presence of
some overhead.
Thread scheduling is done by Julia's built-in {\tt Threads} package.
While parallelizing subproblems across time periods is sufficient to
  solve the presented cases within the time limit,
additional gains may be obtained by decomposing in space as well.
% for larger instances one may need to consider spatial decomposition as well.}
Overall, these results suggest that parallelization can be an effective approach
to balancing solution quality and runtime requirements in industrial-scale AC
Unit Commitment.

\section{Conclusion}
\label{sec:conclusion}

This work provides a simple decomposition framework for solving large-scale instances
of a challenging AC Unit Commitment problem posed by Challenge 3 of the Grid Optimization Competition.
Although the competition formulation includes several features that this work does not consider, such as contingency constraints and line switching,
the proposed algorithms are successful in producing high-quality solutions to the majority of problem instances considered.
A variant of the proposed method using a parallel
decomposition of AC-OPF subproblems is capable of producing solutions to all instances
considered within the two hour time limit, with some reduction in solution quality
compared to a fully sequential computational approach.
% The following removed for space. -RP
%A comparison of different algorithm variations illustrates that it is challenging to balance reserve feasibility
%and AC power flow feasibility, however  effective heuristics that balance these two constraints are possible. 
%All algorithms considered are benchmarked on 28 problem instances for the day-ahead market provided by the Grid Optimization Competition.
The success of this algorithm and those from other participants in the grid optimization
competition suggest that industry-scale AC Unit Commitment is within reach of current
numerical optimization methods.

\section*{Acknowledgements}
This work was funded by the U.S. Department of Energy Advanced Research
Projects Agency-Energy (ARPA-E), as part of the Grid Optimization Competition.
LA-UR-23-31000.

\begin{table*}
  \centering
  \caption{Solution Quality Results for Algorithm \ref{alg:reserve-preserving}}
%\resizebox{\textwidth}{!}{
  \begin{tabular}{|cccrcccr|}
    \hline\hline
    % TODO: Computational cost for best-known solution?
    Case ID & Best-known solution (\$) & Objective (\$) & \multicolumn{1}{c}{Gap} & Res. penalty (\$) & $p$-penalty (\$) & $q$-penalty (\$) & \multicolumn{1}{c}{Solve time (s)} \\
    \hline\hline
N00073-303 & +1.48e+08 & +1.48e+08 &   0.03\% & -4.90e+04 & -2.42e+00 & -3.18e+00 & 181 \\
N00073-333 & +2.31e+08 & +2.30e+08 &   0.02\% & -7.36e+04 & -1.44e+01 & -2.70e+01 & 180 \\
N00073-373 & +2.35e+08 & +2.34e+08 &   0.48\% & -3.43e+05 & -7.28e+01 & -3.91e+02 & 187 \\
\bf N00073-991 & \bf +5.90e+07 & \bf +7.95e+06 &  \bf 86.51\% & \bf -1.16e+05 & \bf -6.50e+01 & \bf -4.92e+07 & \bf 193 \\
\hline
N00617-002 & +1.64e+08 & +1.64e+08 &   0.17\% & -3.58e-12 & -2.68e+02 & -1.45e+03 & 257 \\
N00617-015 & +2.65e+08 & +2.65e+08 &   0.11\% & -2.55e+04 & -3.58e+02 & -2.21e+03 & 304 \\
N00617-039 & +1.64e+08 & +1.64e+08 &   0.18\% & -2.72e+04 & -2.79e+02 & -1.49e+03 & 299 \\
N00617-069 & +2.66e+08 & +2.65e+08 &   0.31\% & -8.14e+03 & -6.91e+02 & -5.04e+03 & 295 \\
\hline
N02000-022 & +7.59e+08 & +7.31e+08 &   3.78\% & -2.45e+07 & -1.27e+02 & -1.08e+03 & 984 \\
N02000-031 & +8.26e+08 & +7.84e+08 &   5.12\% & -1.83e+07 & -1.64e+07 & -1.73e+03 & 2621 \\
\bf N02000-074 & \bf +7.58e+08 & \bf +5.13e+08 &  \bf 32.36\% & \bf -2.05e+06 & \bf -2.44e+08 & \bf -1.97e+03 & \bf 608 \\
\bf N02000-080 & \bf +8.30e+08 & \bf +5.51e+08 &  \bf 33.62\% & \bf -5.83e+05 & \bf -2.79e+08 & \bf -6.09e+03 & \bf 641 \\
\hline
\bf N04224-005 & \bf +4.96e+08 & \bf +2.18e+08 & \bf  55.99\% & \bf -5.01e+07 & \bf -5.56e+07 & \bf -1.52e+08 & \bf 1813 \\
\bf N04224-022 & \bf +4.96e+08 & \bf -5.32e+07 & \bf 110.71\% & \bf -2.59e+07 & \bf -2.01e+08 & \bf -3.04e+08 & \bf 2013 \\
\bf N04224-038 & \bf +4.97e+08 & \bf +2.24e+08 & \bf  54.82\% & \bf -3.11e+07 & \bf -9.72e+07 & \bf -1.26e+08 & \bf 1729 \\
\bf N04224-054 & \bf +7.22e+08 & \bf +4.92e+08 & \bf  31.90\% & \bf -2.26e+08 & \bf -3.72e+05 & \bf -1.22e+05 & \bf 1074 \\
\hline
N06049-003 & +6.09e+08 & +5.71e+08 &   6.30\% & -5.64e+05 & -5.93e+05 & -1.24e+07 & 5173 \\
\bf N06049-031 & \bf +6.78e+08 & \bf +4.09e+08 &  \bf 39.69\% & \bf -1.18e+06 & \bf -2.54e+05 & \bf -2.63e+07 & \bf 4360 \\
\bf N06049-057 & \bf +6.09e+08 & \bf +2.83e+08 &  \bf 53.58\% & \bf -3.47e+05 & \bf -3.19e+05 & \bf -9.33e+07 & \bf 3887 \\
\bf N06049-069 & \bf +8.27e+08 & \bf +3.82e+08 &  \bf 53.74\% & \bf -3.22e+05 & \bf -2.65e+06 & \bf -9.78e+06 & \bf 5095 \\
\hline
N06717-044 & +9.05e+08 & +8.14e+08 &  10.05\% & -2.02e+07 & -3.15e+07 & -1.65e+04 & 3870 \\
N06717-050 & +1.32e+09 & +1.23e+09 &   6.79\% & -1.80e+07 & -1.97e+07 & -3.34e+04 & 3748 \\
\bf N06717-062 & \bf +9.11e+08 & \bf +2.99e+08 &  \bf 67.20\% & \bf -1.33e+07 & \bf -5.50e+08 & \bf -4.41e+06 & \bf 3407 \\
\bf N06717-068 & \bf +1.33e+09 & \bf +3.11e+08 &  \bf 76.57\% & \bf -2.05e+07 & \bf -9.31e+08 & \bf -1.53e+06 & \bf 4242 \\
\hline
N08316-109 & +1.43e+09 & +1.36e+09 &   4.55\% & -1.94e+06 & -5.22e+07 & -4.07e+06 & 4661 \\
N08316-121 & +1.16e+09 & +1.03e+09 &  11.28\% & -3.37e+06 & -2.32e+07 & -9.62e+07 & 5723 \\
N08316-131 & +1.43e+09 & +1.41e+09 &   1.85\% & -6.96e+06 & -5.66e+06 & -5.01e+06 & 4928 \\
\bf N08316-157 & \bf +1.18e+09 & \bf +6.44e+08 &  \bf 45.54\% & \bf -4.01e+06 & \bf -2.59e+08 & \bf -2.66e+08 & \bf 6957 \\
    \hline\hline
  \end{tabular}
%}
  \label{tab:reserve-preserving}
\end{table*}

\begin{table*}
  \centering
  \caption{Solution Quality Results for Algorithm \ref{alg:sequential-greedy}}
%\resizebox{\textwidth}{!}{
  \begin{tabular}{|cccrcccr|}
    \hline\hline
    Case ID & Best-known solution (\$) & Objective (\$) & \multicolumn{1}{c}{Gap} & Res. penalty (\$) & $p$-penalty (\$) & $q$-penalty (\$) & \multicolumn{1}{c}{Solve time (s)} \\
    \hline\hline
N00073-303 & +1.48e+08 & +1.48e+08 &   0.11\% & -9.39e+02 & -2.42e+00 & -3.18e+00 & 163 \\
N00073-333 & +2.31e+08 & +2.30e+08 &   0.11\% & -1.94e+04 & -1.44e+01 & -2.70e+01 & 165 \\
N00073-373 & +2.35e+08 & +2.34e+08 &   0.48\% & -2.00e+05 & -7.28e+01 & -3.91e+02 & 171 \\
\bf N00073-991 & \bf +5.90e+07 & \bf +7.96e+06 &  \bf 86.50\% & \bf -9.12e+04 & \bf -6.50e+01 & \bf -4.92e+07 & \bf 188 \\
\hline
N00617-002 & +1.64e+08 & +1.64e+08 &   0.01\% & -0.00e+00 & -5.39e+01 & -2.69e+02 & 253 \\
N00617-015 & +2.65e+08 & +2.65e+08 &   0.12\% & -2.70e+05 & -2.29e+02 & -1.22e+03 & 271 \\
N00617-039 & +1.64e+08 & +1.64e+08 &   0.17\% & -2.70e+05 & -7.80e+01 & -4.14e+02 & 262 \\
N00617-069 & +2.66e+08 & +2.66e+08 &   0.10\% & -2.70e+05 & -2.46e+02 & -1.30e+03 & 280 \\
\hline
\bf N02000-022 & \bf +7.59e+08 & \bf -2.01e+09 & \bf 364.29\% & \bf -2.77e+09 & \bf -6.88e+01 & \bf -3.14e+02 & \bf 1345 \\
\bf N02000-031 & \bf +8.26e+08 & \bf -4.40e+09 & \bf 632.79\% & \bf -5.23e+09 & \bf -8.40e+01 & \bf -4.31e+02 & \bf 2886 \\
\bf N02000-074 & \bf +7.58e+08 & \bf +1.62e+08 & \bf  78.64\% & \bf -5.96e+08 & \bf -5.18e+01 & \bf -2.89e+02 & \bf 926 \\
\bf N02000-080 & \bf +8.30e+08 & \bf -1.87e+07 & \bf 102.25\% & \bf -8.49e+08 & \bf -6.19e+01 & \bf -2.93e+02 & \bf 943 \\
\hline
\bf N04224-005 & \bf +4.96e+08 & \bf +4.40e+08 &  \bf 11.34\% & \bf -2.78e+05 & \bf -2.08e+07 & \bf -3.01e+07 & \bf 2240 \\
\bf N04224-022 & \bf +4.96e+08 & \bf +2.92e+08 &  \bf 41.26\% & \bf -0.00e+00 & \bf -8.21e+07 & \bf -1.17e+08 & \bf 2246 \\
\bf N04224-038 & \bf +4.97e+08 & \bf +2.57e+08 &  \bf 48.20\% & \bf -0.00e+00 & \bf -9.04e+07 & \bf -1.41e+08 & \bf 2068 \\
\bf N04224-054 & \bf +7.22e+08 & \bf +6.05e+08 &  \bf 16.22\% & \bf -1.15e+08 & \bf -1.82e+03 & \bf -9.52e+03 & \bf 2471 \\
\hline
N06049-003 & +6.09e+08 & +5.95e+08 &   2.27\% & -6.95e+05 & -8.79e+04 & -1.00e+07 & 5951 \\
\bf N06049-031 & \bf +6.78e+08 & \bf +5.54e+08 &  \bf 18.32\% & \bf -1.13e+07 & \bf -1.08e+05 & \bf -1.61e+07 & \bf 5283 \\
\bf N06049-057 & \bf +6.09e+08 & \bf +3.20e+08 &  \bf 47.57\% & \bf -1.00e+06 & \bf -2.98e+05 & \bf -7.96e+07 & \bf 6050 \\
\bf N06049-069 & \bf +8.27e+08 & \bf +4.40e+08 &  \bf 46.72\% & \bf -1.26e+06 & \bf -1.29e+06 & \bf -5.07e+06 & \bf 7597 \\
\hline
\bf N06717-044 & \bf +9.05e+08 & \bf -4.23e+08 & \bf 146.70\% & \bf -1.33e+09 & \bf -1.42e+02 & \bf -3.82e+02 & \bf 6385 \\
\bf N06717-050 & \bf +1.32e+09 & \bf -1.29e+08 & \bf 109.81\% & \bf -1.45e+09 & \bf -2.92e+02 & \bf -1.65e+03 & \bf 7302 \\
\bf N06717-062 & \bf +9.11e+08 & \bf -5.57e+08 & \bf 161.19\% & \bf -1.47e+09 & \bf -1.55e+02 & \bf -9.73e+02 & \bf 6585 \\
\bf N06717-068 & \bf +1.33e+09 & \bf -4.71e+08 & \bf 135.40\% & \bf -1.80e+09 & \bf -4.77e+02 & \bf -3.40e+03 & \bf 7614 \\
\hline
N08316-109 & +1.43e+09 & +1.42e+09 &   0.76\% & -7.80e+05 & -1.13e+06 & -8.53e+05 & 8866 \\
N08316-121 & +1.16e+09 & +1.06e+09 &   9.09\% & -3.81e+06 & -1.10e+07 & -7.68e+07 & 9318 \\
N08316-131 & +1.43e+09 & +1.42e+09 &   1.23\% & -7.64e+06 & -9.54e+05 & -4.55e+05 & 8838 \\
N08316-157 & +1.18e+09 & +1.07e+09 &   9.57\% & -3.23e+06 & -1.06e+07 & -8.26e+07 & 10000 \\
    \hline\hline
  \end{tabular}
%}
  \label{tab:sequential-greedy}
\end{table*}

\begin{table*}
  \centering
  \caption{Solution Quality Results for Algorithm \ref{alg:balancing}}
%\resizebox{\textwidth}{!}{
  \begin{tabular}{|cccrcccr|}
    \hline\hline
    Case ID & Best-known solution (\$) & Objective (\$) & \multicolumn{1}{c}{Gap} & Res. penalty (\$) & $p$-penalty (\$) & $q$-penalty (\$) & \multicolumn{1}{c}{Solve time (s)} \\
    \hline\hline
N00073-303 & +1.48e+08 & +1.48e+08 &   0.09\% & -2.11e+03 & -6.09e+01 & -3.73e+02 & 217 \\
N00073-333 & +2.31e+08 & +2.30e+08 &   0.09\% & -1.65e+04 & -4.70e+01 & -2.87e+02 & 212 \\
N00073-373 & +2.35e+08 & +2.34e+08 &   0.39\% & -2.69e+05 & -1.32e+02 & -7.99e+02 & 212 \\
N00073-991 & +5.90e+07 & +7.36e+06 &  87.52\% & -1.12e+05 & -3.51e+02 & -5.05e+07 & 239 \\
\hline
N00617-002 & +1.64e+08 & +1.64e+08 &   0.09\% & -3.64e-03 & -6.19e+01 & -2.64e+02 & 342 \\
N00617-015 & +2.65e+08 & +2.65e+08 &   0.12\% & -1.32e+03 & -4.25e+02 & -8.20e+02 & 368 \\
N00617-039 & +1.64e+08 & +1.64e+08 &   0.10\% & -1.26e+04 & -5.03e+01 & -2.24e+02 & 337 \\
N00617-069 & +2.66e+08 & +2.66e+08 &   0.09\% & -1.39e+03 & -1.89e+02 & -3.91e+02 & 380 \\
\hline
N02000-022 & +7.59e+08 & +7.47e+08 &   1.61\% & -1.21e+07 & -1.57e+03 & -5.73e+03 & 1614 \\
N02000-031 & +8.26e+08 & +7.29e+08 &  11.73\% & -1.00e+08 & -2.13e+03 & -8.05e+03 & 3154 \\
N02000-074 & +7.58e+08 & +7.57e+08 &   0.08\% & -1.62e+01 & -6.37e+02 & -2.80e+03 & 1831 \\
N02000-080 & +8.30e+08 & +8.29e+08 &   0.08\% & -1.50e+01 & -4.47e+02 & -2.45e+03 & 1763 \\
\hline
N04224-005 & +4.96e+08 & +4.88e+08 &   1.57\% & -1.67e+00 & -9.62e+03 & -2.38e+06 & 1462 \\
N04224-022 & +4.96e+08 & +3.85e+08 &  22.47\% & -1.19e+00 & -4.38e+07 & -6.12e+07 & 1454 \\
N04224-038 & +4.97e+08 & +3.19e+08 &  35.77\% & -6.00e-09 & -6.66e+07 & -1.05e+08 & 1464 \\
N04224-054 & +7.22e+08 & +7.19e+08 &   0.43\% & -4.63e-02 & -2.60e+03 & -2.59e+04 & 902 \\
\hline
N06049-003 & +6.09e+08 & +5.75e+08 &   5.57\% & -2.00e+05 & -1.69e+05 & -1.36e+07 & 4872 \\
N06049-031 & +6.78e+08 & +5.38e+08 &  20.65\% & -1.89e+05 & -1.56e+05 & -1.78e+07 & 4581 \\
N06049-057 & +6.09e+08 & +2.92e+08 &  52.17\% & -1.37e+03 & -3.49e+05 & -8.46e+07 & 3937 \\
N06049-069 & +8.27e+08 & +3.84e+08 &  53.58\% & -1.40e-10 & -1.16e+06 & -8.18e+06 & 4532 \\
\hline
N06717-044 & +9.05e+08 & +8.94e+08 &   1.26\% & -1.28e+07 & -8.45e+02 & -3.59e+03 & 4799 \\
N06717-050 & +1.32e+09 & +1.32e+09 &   0.22\% & -2.41e+06 & -8.21e+02 & -5.60e+03 & 5008 \\
N06717-062 & +9.11e+08 & +8.31e+08 &   8.79\% & -1.34e+07 & -6.33e+07 & -2.45e+05 & 4503 \\
N06717-068 & +1.33e+09 & +1.23e+09 &   7.20\% & -9.18e+06 & -8.07e+07 & -7.49e+04 & 5313 \\
\hline
N08316-109 & +1.43e+09 & +1.41e+09 &   1.29\% & -3.04e+05 & -5.99e+06 & -5.25e+06 & 6938 \\
N08316-121 & +1.16e+09 & +1.03e+09 &  11.02\% & -2.09e+06 & -1.97e+07 & -9.93e+07 & 6469 \\
\bf N08316-131 & \bf +1.43e+09 & \bf +1.40e+09 &   \bf 1.96\% & \bf -9.57e+06 & \bf -5.87e+06 & \bf -5.61e+06 & \bf 8061 \\
N08316-157 & +1.18e+09 & +1.06e+09 &  10.60\% & -1.26e+06 & -9.71e+06 & -1.10e+08 & 6747 \\
    \hline\hline
  \end{tabular}
%}
  \label{tab:balancing}
\end{table*}

\begin{table*}
  \centering
  \caption{Solution Quality Results for Algorithm \ref{alg:parallel}}
%\resizebox{\textwidth}{!}{
  \begin{tabular}{|cccrcccr|}
    \hline\hline
    Case ID & Best-known solution (\$) & Objective (\$) & \multicolumn{1}{c}{Gap} & Res. penalty (\$) & $p$-penalty (\$) & $q$-penalty (\$) & \multicolumn{1}{c}{Solve time (s)} \\
    \hline\hline
N00073-303 & +1.48e+08 & +1.48e+08 &   0.09\% & -2.11e+03 & -6.09e+01 & -3.73e+02 & 215 \\
N00073-333 & +2.31e+08 & +2.30e+08 &   0.09\% & -1.65e+04 & -4.70e+01 & -2.87e+02 & 207 \\
N00073-373 & +2.35e+08 & +2.34e+08 &   0.39\% & -2.69e+05 & -1.32e+02 & -7.99e+02 & 209 \\
N00073-991 & +5.90e+07 & +7.36e+06 &  87.52\% & -1.12e+05 & -3.51e+02 & -5.05e+07 & 234 \\
\hline
N00617-002 & +1.64e+08 & +1.64e+08 &   0.09\% & -3.64e-03 & -6.19e+01 & -2.64e+02 & 249 \\
N00617-015 & +2.65e+08 & +2.65e+08 &   0.12\% & -1.32e+03 & -4.25e+02 & -8.20e+02 & 273 \\
N00617-039 & +1.64e+08 & +1.64e+08 &   0.10\% & -1.26e+04 & -5.03e+01 & -2.24e+02 & 265 \\
N00617-069 & +2.66e+08 & +2.66e+08 &   0.09\% & -1.39e+03 & -1.89e+02 & -3.91e+02 & 296 \\
\hline
N02000-022 & +7.59e+08 & +7.47e+08 &   1.61\% & -1.21e+07 & -1.58e+03 & -5.90e+03 & 755 \\
N02000-031 & +8.26e+08 & +7.29e+08 &  11.73\% & -1.00e+08 & -1.98e+03 & -7.70e+03 & 2390 \\
N02000-074 & +7.58e+08 & +7.57e+08 &   0.08\% & -1.59e+01 & -4.11e+02 & -2.30e+03 & 738 \\
N02000-080 & +8.30e+08 & +8.29e+08 &   0.08\% & -1.55e+01 & -3.61e+02 & -2.05e+03 & 652 \\
\hline
N04224-005 & +4.96e+08 & +1.80e+08 &  63.61\% & -6.81e-01 & -2.19e+08 & -9.09e+07 & 827 \\
N04224-022 & +4.96e+08 & +2.74e+08 &  44.84\% & -9.18e-01 & -1.61e+08 & -5.56e+07 & 775 \\
N04224-038 & +4.97e+08 & +2.06e+08 &  58.53\% & -4.14e-09 & -2.01e+08 & -8.42e+07 & 921 \\
N04224-054 & +7.22e+08 & +7.19e+08 &   0.47\% & -4.63e-02 & -3.32e+05 & -2.58e+04 & 720 \\
\hline
N06049-003 & +6.09e+08 & +5.64e+08 &   7.39\% & -2.03e+05 & -1.12e+07 & -1.32e+07 & 1934 \\
N06049-031 & +6.78e+08 & +5.38e+08 &  20.68\% & -1.88e+05 & -4.29e+05 & -1.76e+07 & 2059 \\
N06049-057 & +6.09e+08 & +2.92e+08 &  52.06\% & -1.59e+03 & -3.47e+05 & -8.36e+07 & 1537 \\
N06049-069 & +8.27e+08 & +3.80e+08 &  53.97\% & -1.57e-10 & -2.16e+06 & -1.01e+07 & 2636 \\
\hline
N06717-044 & +9.05e+08 & +8.94e+08 &   1.26\% & -1.28e+07 & -8.09e+02 & -3.50e+03 & 3868 \\
N06717-050 & +1.32e+09 & +1.32e+09 &   0.22\% & -2.41e+06 & -8.65e+02 & -5.65e+03 & 3977 \\
N06717-062 & +9.11e+08 & +8.31e+08 &   8.79\% & -1.34e+07 & -6.33e+07 & -2.44e+05 & 3217 \\
N06717-068 & +1.33e+09 & +1.23e+09 &   7.20\% & -9.18e+06 & -8.07e+07 & -7.49e+04 & 4379 \\
\hline
N08316-109 & +1.43e+09 & +1.38e+09 &   3.14\% & -3.05e+05 & -3.23e+07 & -5.18e+06 & 4114 \\
N08316-121 & +1.16e+09 & +1.01e+09 &  13.00\% & -2.10e+06 & -3.78e+07 & -1.05e+08 & 3046 \\
N08316-131 & +1.43e+09 & +1.39e+09 &   3.24\% & -9.74e+06 & -2.41e+07 & -5.58e+06 & 3995 \\
N08316-157 & +1.18e+09 & +1.02e+09 &  13.59\% & -1.20e+06 & -4.37e+07 & -1.12e+08 & 3438 \\
    \hline\hline
  \end{tabular}
%}
  \label{tab:parallel}
\end{table*}

\bibliographystyle{IEEEtran}
\bibliography{IEEEabrv,ref}
%\begin{thebibliography}{1}
%\bibitem{Shell}
%M.~Shell, \emph{How to Use the IEEEtran Latex Class}, Latex Archive Contents, \verb+http://www.ieee.org/conferences_events/+ \verb+conferences/publishing/templates.htm+
%\end{thebibliography}

\end{document}